\documentclass[usenatbib,useAMS]{mnras}
\usepackage{graphicx}
\usepackage{amssymb}

\usepackage{newtxtext,newtxmath}
\usepackage[T1]{fontenc}
\usepackage{ae,aecompl}

\title [Disc evolution in binaries]
{   
The evolution of photo-evaporating viscous discs in binaries
}
\author[G. P. Rosotti and C. J. Clarke]{
Giovanni P. Rosotti$^{1}$\thanks{E-mail: rosotti@ast.cam.ac.uk} and
Cathie J. Clarke$^{1}$
\\
$^1$Institute of Astronomy, Madingley Rd, Cambridge, CB3 0HA, UK \\} 

\date{Accepted 2017 October 23. Received 2017 October 14; in original form 2017 April 13}

\pubyear{2017}

\begin {document}

\label{firstpage}
\pagerange{\pageref{firstpage}--\pageref{lastpage}} 

\maketitle

\begin{abstract}
A large fraction of stars are in binary systems, yet the evolution of proto-planetary discs in binaries has been little explored from the theoretical side. In this paper we investigate the evolution of the discs surrounding the primary and secondary components of binary systems on the assumption that this is driven by photoevaporation induced by X-rays from the respective star. We show how for close enough separations (20-30 AU for average X-ray luminosities) the tidal torque of the companion changes the qualitative behaviour of disc dispersal from inside out to outside in. Fewer transition discs created by photoevaporation are thus expected in binaries. We also demonstrate that in close binaries the reduction in viscous time leads to accelerated disc clearing around both components, consistent with \textit{unresolved} observations. When looking at the \textit{differential} disc evolution around the two components, in close binaries discs around the secondary clear first due to the shorter viscous timescale associated with the smaller outer radius. In wide binaries instead the difference in photo-evaporation rate makes the secondaries longer lived, though this is somewhat dependent on the assumed scaling of viscosity with stellar mass. We find that our models are broadly compatible with the growing sample of \textit{resolved} observations of discs in binaries. We also predict that binaries have higher accretion rates than single stars for the same disc mass. Thus binaries probably contribute to the observed scatter in the relationship between disc mass and accretion rate in young stars.
\end{abstract}


\section{Introduction}


A large fraction of stars are in binary systems \citep{Raghavan2010}. Despite the possible impediments to planet formation in binaries \citep{Thebault2011,Rafikov2013,Marzari2013,Lines2015}, the growing
census of planets discovered in binary systems attests to the fact that
planet formation is indeed viable in these environments, either around one of the two components \citep[e.g.,][]{Hatzes2003,Dumusque2012} or around the binary itself \citep{Doyle2011,Welsh2012}. 
In addition,  planet formation is likely to be affected
by the influence of binarity on disc lifetime and this has spurred a number of
observational efforts to characterise the disc bearing properties of stars in young
binaries \citep[e.g.,][]{Cieza2009,Kraus2011,Daemgen2012,Daemgen2013}.

Such surveys offer a broader opportunity to test our understanding of protoplanetary
disc evolution. Disc evolution is widely modelled \citep{Clarke2001,Alexander2006,Owen2011} in terms of parametrised viscous
evolution which is terminated when such evolution becomes dominated by photoevaporation
by the central star. Such models, as applied to single stars,
have been shown to be compatible with a wide
range of observed diagnostics, such as the evolution of the disc fraction with time \citep[e.g.,][]{Haisch2001,Fedele2010,Ribas2014}, and naturally provide an explanation for at least some transition discs \citep{OwenClarke2012}, i.e. discs that show evidence for an inner hole; such models have not however been tested in binary systems. In this paper we test whether such
models are consistent with the available data on discs in binary systems, drawing
both on resolved observations (where the presence of discs around individual
components is detected, e.g. \citealt{Daemgen2012,Daemgen2013}) and unresolved observations (which instead indicate the
presence of a disc in at least one member of the pair, e.g. \citealt{Cieza2009,Kraus2011}).


There are two distinct differences that apply to discs in binaries, which can be used to our advantage. Firstly, binary components are coeval to an
excellent approximation. Theoretically, the only viable scenario for binary formation is the fragmentation of a molecular cloud core (see \citealt{Reipurth2014} for a recent review on the subject), which implies a difference in age of up to $\sim 10^5$ years (the free fall time of the natal core), a small fraction of
the typical ages of T Tauri stars. This coevality has been
confirmed in a number of systems by placing components on pre-main sequence tracks
\citep{KrausHillenbrand2009}: indeed \citet{Daemgen2012} confirmed
coevality   
in all cases where low veiling
permits an accurate age calibration (see also the review of \citealt{Stassun2014}). 
This is in contrast to the situation within star formation regions
where age spreads may be large and are poorly quantified
(see e.g. \citealt{PallaStahler2000,Hillenbrand2008,Jeffries2011}). The coeval nature of binary components then makes it possible to study disc evolution as a function of stellar mass.
Secondly, the disc outer radius is well constrained in binary systems by the
tidal effect of the companion star: this imposes a zero mass flux outer
boundary to the disc at a radius that is a dynamically  determined
function of binary mass ratio and separation (\citealt{PapaloizouPringle1977,Pichardo2005}; see also the observational study of \citealt{Harris2012}).
In discs whose evolution is driven by accretion on to the star, the evolution
timescale at all radii in the disc is given by the viscous timescale at the
disc's outer edge \citep{Pringle81}, and it is a significant disadvantage that this
quantity is not well constrained observationally in single stars. The well defined outer radius of a binary provides a laboratory to study the dependence of the viscous timescale on radius.



In this paper we assume that disc dispersal is driven by X-ray photoevaporation.
This is motivated by the fact that the dependence of X-ray luminosity
(and hence photoevaporation rate) is well quantified as a function of stellar
mass (see \citealt{Preibisch2005,Guedel2007}) which is a necessary ingredient
when examining differential disc lifetimes in binaries. Modelling photoevaporation by
EUV radiation is complicated by the fact
that the EUV luminosity of T T Tauri stars (and its mass dependence) is not
well constrained observationally \citep{Alexander2005}, while FUV photoevaporation \citep{Gorti2009} by the central
star has not yet been subject to radiation hydrodynamical modelling. In the X-ray
case, mass loss rates vary roughly linearly with X-ray luminosity and, as noted above,
give rise to models of single star disc evolution that
are broadly compatible with observations.

This paper is organised as follows. Section \ref{sec:model} introduces the model we use to describe disc evolution and section \ref{sec:results} presents our results about disc evolution. We then compare our results with the observations in section \ref{sec:discussion} and we finally draw our conclusions in section \ref{sec:conclusions}.

\section{Description of the model}
\label{sec:model}
 We assume that the discs surrounding each star within a binary system
evolve independently. Strictly speaking this means that our calculation
is only applicable to the phase of disc evolution when re-supply of gas
from beyond the binary orbit can be neglected. Empirically, this is motivated by the absence of substantial circumbinary discs in most binaries that are wider than a few A.U. (e.g., \citealt{Jensen96}; see also the discussion in \citealt{Monin2007}).


  We also assume that the photoevaporation of each disc is dominated by the
wind driven by the X-ray luminosity of its respective central star, i.e.
that there is no cross-over of X-ray heating between discs. This is likely
to be a good approximation for binaries whose separation exceeds the
radius in the disc ($R_X$) within which X-ray photoevaporation is effective:
 $R_X$ is about $80$ A.U. for a solar mass star and scales linearly with
stellar mass \citep{Owen2012}. For closer binaries there is some possibility of cross-over,
although this cannot be quantified without three-dimensional simulations.
Nevertheless, this omission is unlikely to affect any of our conclusions:
it will only be significant in the case of close, extreme mass ratio binaries,
in which case the extra flux from the X-ray luminous primary may
accelerate the clearing of the secondary's disc. This is however a regime
in which the secondary's disc in any case clears significantly faster
than the primary's by purely viscous processes.

\subsection{Method}

\begin{figure}
\includegraphics[width=\columnwidth]{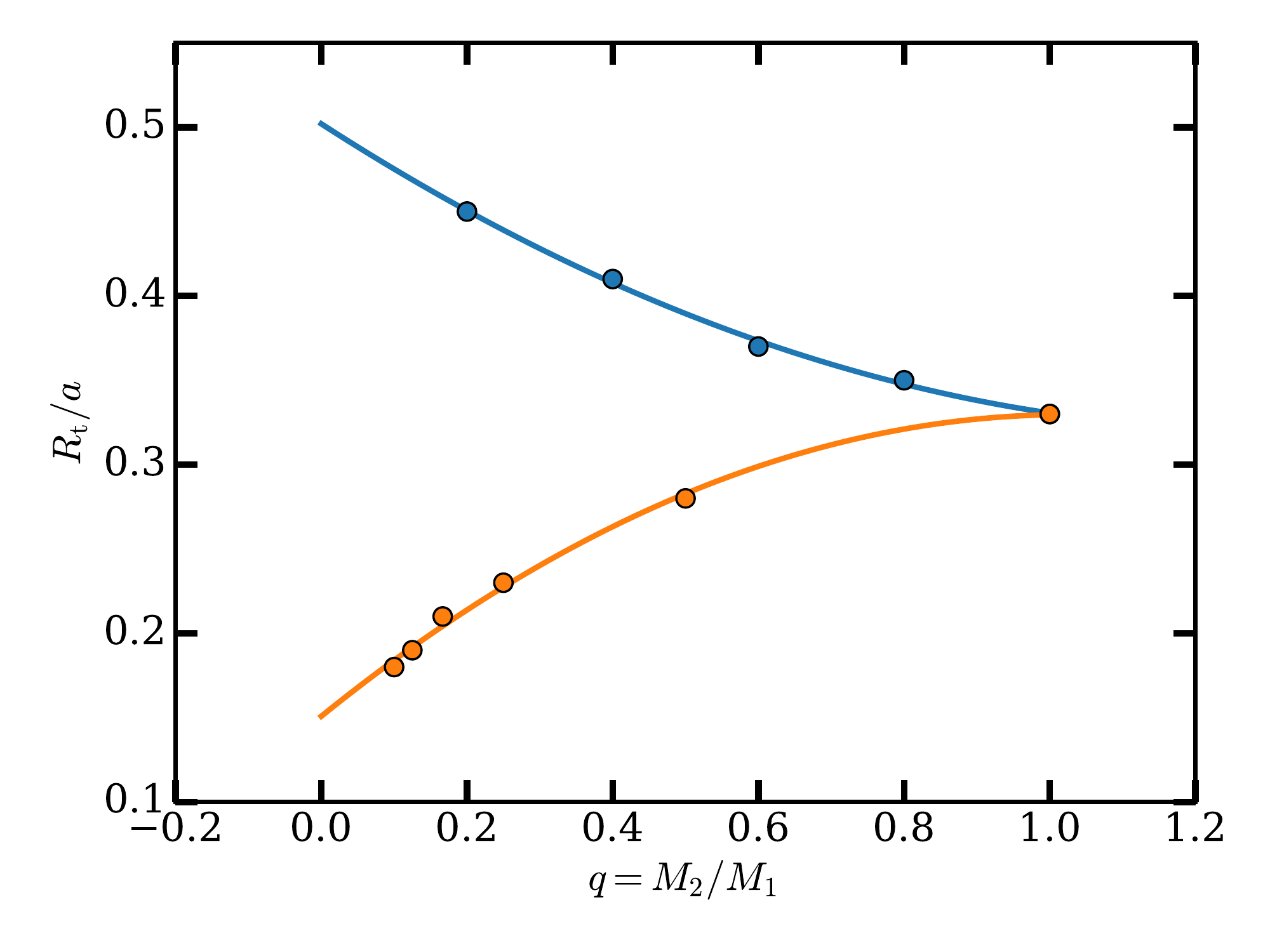}
\caption{The tidal radii of the primary/secondary discs
(upper and lower curves) normalised to the binary separation as
a function of binary mass ratio, $q=M_2/M_1$. Datapoints from
\citet{PapaloizouPringle1977}, while the curves are the numerical fits
employed in this work.}
\label{fig:truncation}
\end{figure}

  For a binary of given component masses ($M_1$ and $M_2$),mass
ratio $q$ ($ = M_2/M_1$) and separation, $a$, the tidal truncation
radius, $R_\mathrm{t}$, of each disc is calculated using the formula given in
\citet{PapaloizouPringle1977} (see Figure \ref{fig:truncation}). Note that for a mass ratio of unity this produces the well-known result that the truncation radius $R_\mathrm{t} \simeq a/3$. The initial mass of each disc is assigned
a value equal to $0.1 \times$ the mass of its parent star; this is
distributed with a surface density profile:

\begin{equation}
\Sigma(R,0) = {{C}\over{R}} \rm{exp} \biggl(-{{R}\over{R_1}}\biggr)
\label{eq:self_similar}
\end{equation}

 where $C$ is adjusted to give the correct disc mass within a 
radius $R_\mathrm{t}$\footnote{We have checked that the results are not significantly changed if $C$ is adjusted so that 10 percent of the stellar mass instead corresponds to the total mass the disc would have out to infinite radius.}. Our initial conditions correspond to a viscous similarity solution 
\citep{LyndenBellPringle74,Hartmann98} for
a freely expanding disc with kinematic viscosity $\nu \propto R$
(see below) and was the functional form adopted by
\citet{Owen2010} (with $R_1 = 18$ A.U., {which we assume in this work}) as a plausible initial
condition that  matched the resulting model properties 
to observed discs in single stars. Note that this initial
distribution (and its dependence on stellar mass) is poorly constrained 
observationally. The $\Sigma \propto R^{-1}$ dependence
(for $R < R_1$) is motivated by the observed power law decline in
disc surface density inferred 
from mm imaging \citep{WilliamsCieza2011}; in the absence of information
about the scaling of viscosity with stellar mass (cf. \citealt{AlexanderArmitage2006,Dullemond2006}) we set the viscosity law as 
$\nu = \nu_0 (R/R_0)$ where the value $\nu_0/R_0 = 10^{-5}$ A.U. yr$^{-1}$
yields observationally reasonable disc lifetimes of order
a few Myr \citep{Haisch2001,Fedele2010}.

 X-ray photoevaporation is included by applying the parametrisation of
X-ray mass loss per unit area ($\dot \Sigma_X$)
given in \citet{Owen2012} 
The mass-loss profile has the property that the radial scaling is proportional to stellar
mass, that is,
\begin{equation}
\dot \Sigma_X=\dot \Sigma_X (R M_\ast^{-1})
\label{eq:radial_scaling}
\end{equation}

while the total wind loss
mass rate scales linearly with the X-ray luminosity.
The X-ray luminosity of T Tauri stars has been well characterised by COUP (the Chandra Orion Ultra-deep Project (\citealt{Preibisch2005}; see also \citealt{Guedel2007} for similar results obtained using XMM in Taurus). In Section \ref{sec:results} we assign X-ray luminosities to stellar mass using the
mass dependence of the {\it mean}  X-ray luminosity
i.e. $L_X \propto M_*^{1.44}$;  we note however that  the scatter is large
(standard deviation, $\sigma$, in log$_{10} (L_X)$ of $0.65$) at all masses and in Section \ref{sec:discussion} we
undertake population synthesis in which we select X-ray luminosities
from a mass dependent luminosity function with this $\sigma$.



  The model discs are evolved through integration of the
viscous diffusion equation:

\begin{equation}
{{\partial \Sigma}\over{\partial t}} = {{1}\over{R}} {{\partial}\over{\partial R}} \biggl[ 3 R^{1/2} {{\partial}\over{\partial R}} (\nu \Sigma R^{1/2})\biggr]
-\dot\Sigma_X
\label{eq:sigma_t}
\end{equation}

and the presence of the companion is modelled by imposing a zero flux boundary condition (${{\partial (\nu \Sigma R^{1/2})}/{\partial R}} = 0$) at $R=R_\mathrm{t}$, {as commonly employed in the context of evolved binaries such as dwarf novae (e.g., \citealt{1981MNRAS.194..967B}) and justified by the fact that the tidal torque from the companion has a very steep dependence on radius. In practice, this means that its action is restricted to a very narrow region and our approach is equivalent to assuming that the affected region is infinitesimally thin}. This is often called a closed boundary condition in the context of the theory of differential equations. Equation \ref{eq:sigma_t} is differenced on a grid equispaced in
$R^{1/2}$ and the equation is integrated via a standard finite-difference method, first order accurate in time and second order accurate in space \citep{Pringle86}; the code is described in detail in \citet{ErcolanoRosotti2015}.

\section{Results: the evolution of photoevaporating viscous discs in binaries}
\label{sec:results}
\subsection{The effect of tidal truncation on disc evolution}
\label{sec:mean}

In this section we address how the evolution of a \textit{single} disc around one of the two components of the binary is affected by the closed boundary imposed by the tidal torque of the companion. 

\subsubsection{Qualitative behaviour}

\label{sec:qualitative}
In the case of discs around single stars, the
interplay between photoevaporation and viscous evolution produces
a well defined evolutionary sequence \citep{Clarke2001,Owen2010}: (a) A prolonged phase of viscous
draining that is little modified by the photoevaporative mass loss
(b) a phase of so-called `photoevaporation starved accretion' \citep{Drake2009} where the disc profile becomes somewhat depleted in the region
of maximum mass loss (i.e. at $10$s of AU) (c) The
opening of a gap in the disc at a point where the mass loss rate starts
to decline steeply (i.e. at a few AU) (d) The viscous draining of the
inner disc and (e) The progressive erosion of the outer disc by
photoevaporation. During the last phase (e) (and possibly during phase (d) ), see \citealt{AlexanderArmitage2007}), the disc has an inner hole and is therefore a transition disc.

 In the binary case, phase (a) is modified after significant
material has diffused out to interact with the  tidal boundary condition
at $R_\mathrm{t}$. From this point (in the absence of photoevaporation)
the disc would evolve towards a similarity solution which differs
from that which applies in the case of a freely expanding disc \citep{LyndenBellPringle74}:

\begin{equation}
\Sigma(R,t) = \Sigma_0  \frac{R_1}{R} (1+t/t_{\nu,1})^{-3/2} \rm{exp} \biggl(-{{R}\over{R_1 (1+t/t_{\nu,1})}}\biggr).
\label{eq:self_similar_t}
\end{equation}

{The solution of eq. \ref{eq:sigma_t} with a closed outer boundary and no photo-evaporation can be obtained via separation of variables and is instead}:

\begin{equation}
\Sigma(R,t) = {{A}\over{R^{3/2}}} \rm{sin} \biggl({{\pi  R^{1/2}}\over{2R_\mathrm{t}^{1/2}}}\biggr)\rm exp\biggl(-{{t}\over{t_t}}\biggr),
\label{eq:closed_boundary_sigma}
\end{equation}

{as can be verified by substitution into eq. \ref{eq:sigma_t},} where $t_t = 16 R_\mathrm{t}^2/3 \pi^2 \nu_t$, $\nu_t$ is the kinematic
viscosity at $R_\mathrm{t}$ (i.e. $\nu_0 R_\mathrm{t}/R_0$) {and $A$ and $\Sigma_0$ are normalization constants}.
 Note that, while in both similarity solutions $ \Sigma \propto R^{-1}$ at small
radii, the closed boundary condition leads to a power law
of index $-1.5$ in the outer regions of the disc.
 
\begin{figure}
\includegraphics[width=\columnwidth]{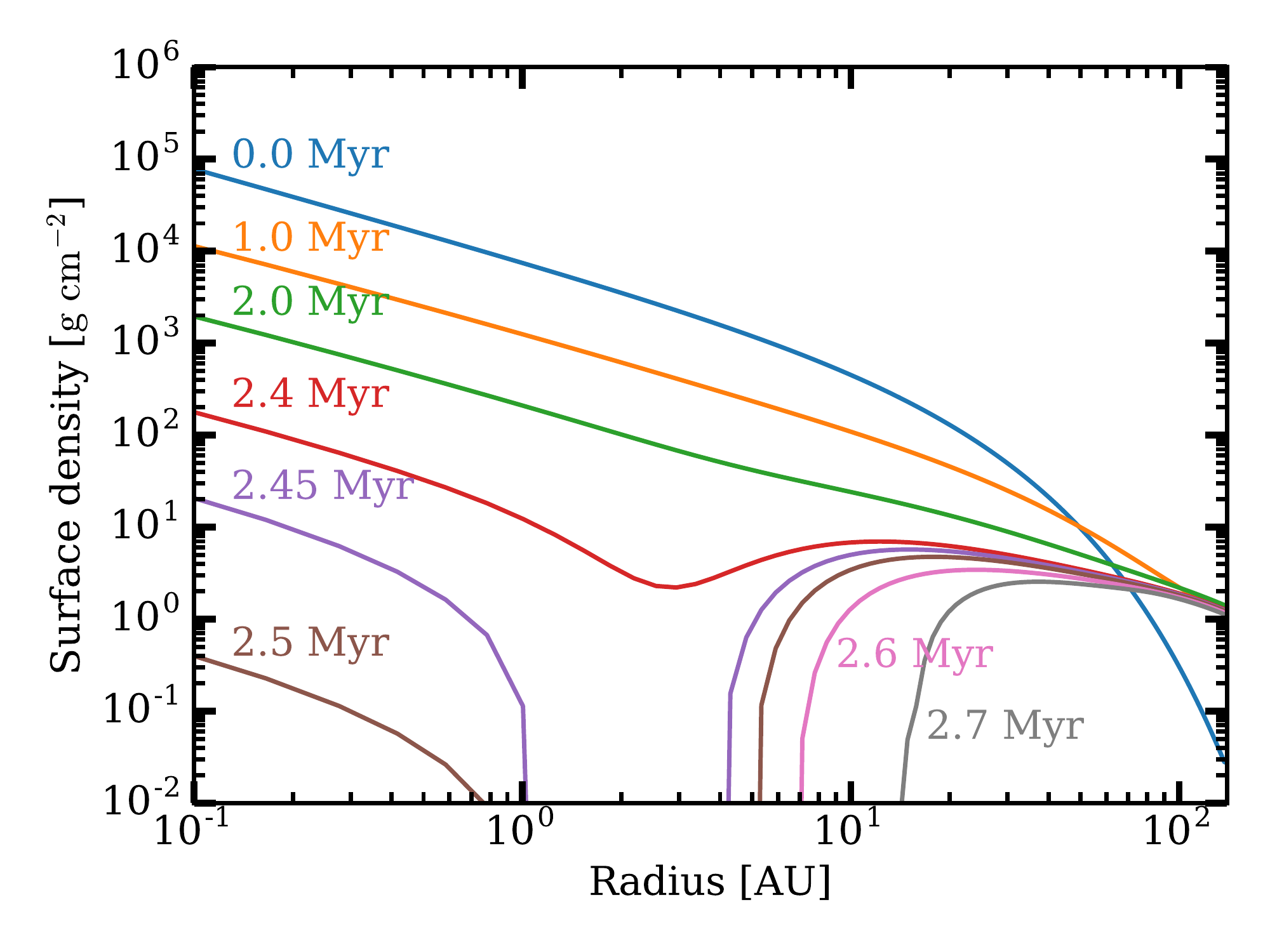}
\includegraphics[width=\columnwidth]{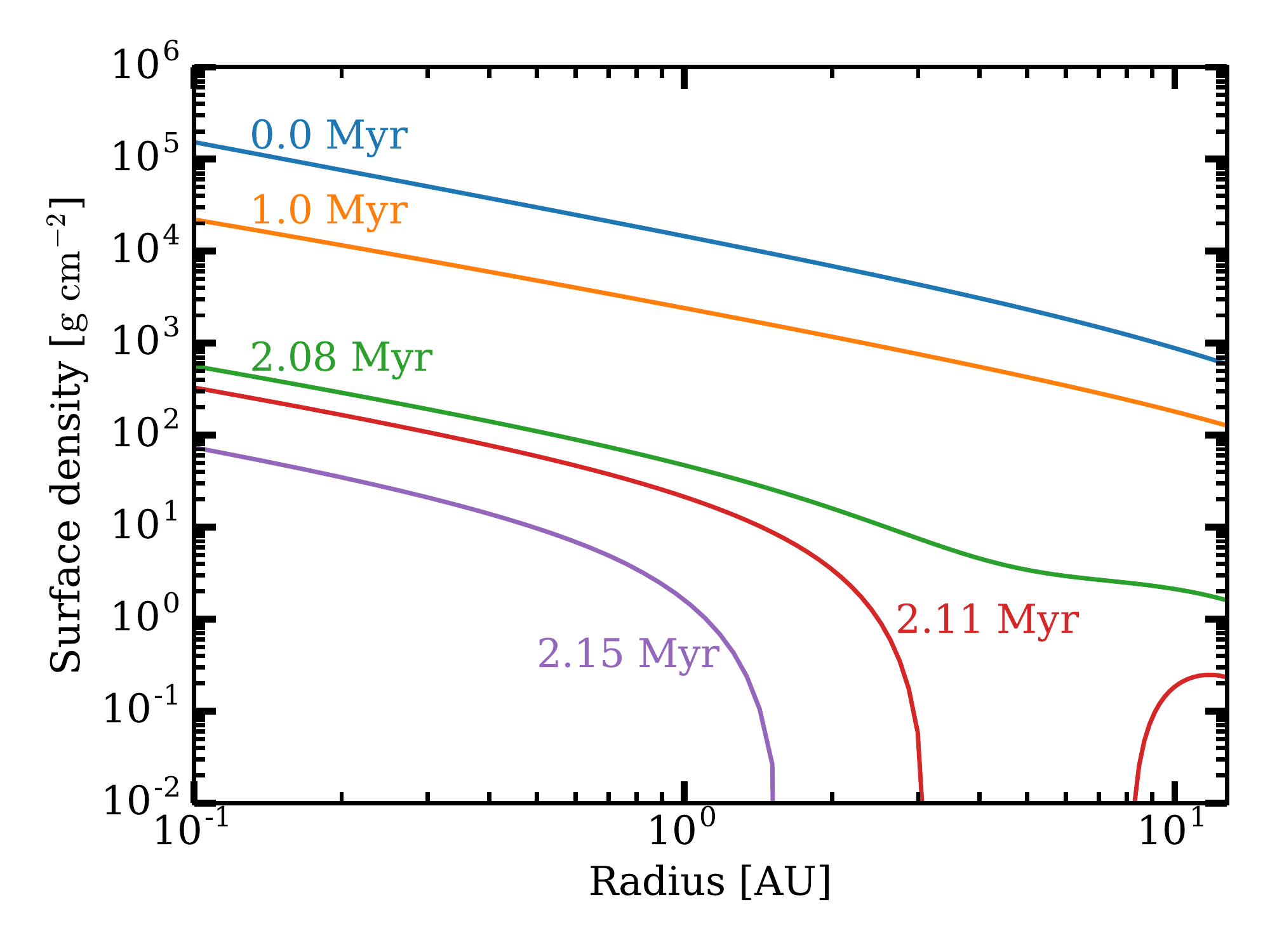}
\includegraphics[width=\columnwidth]{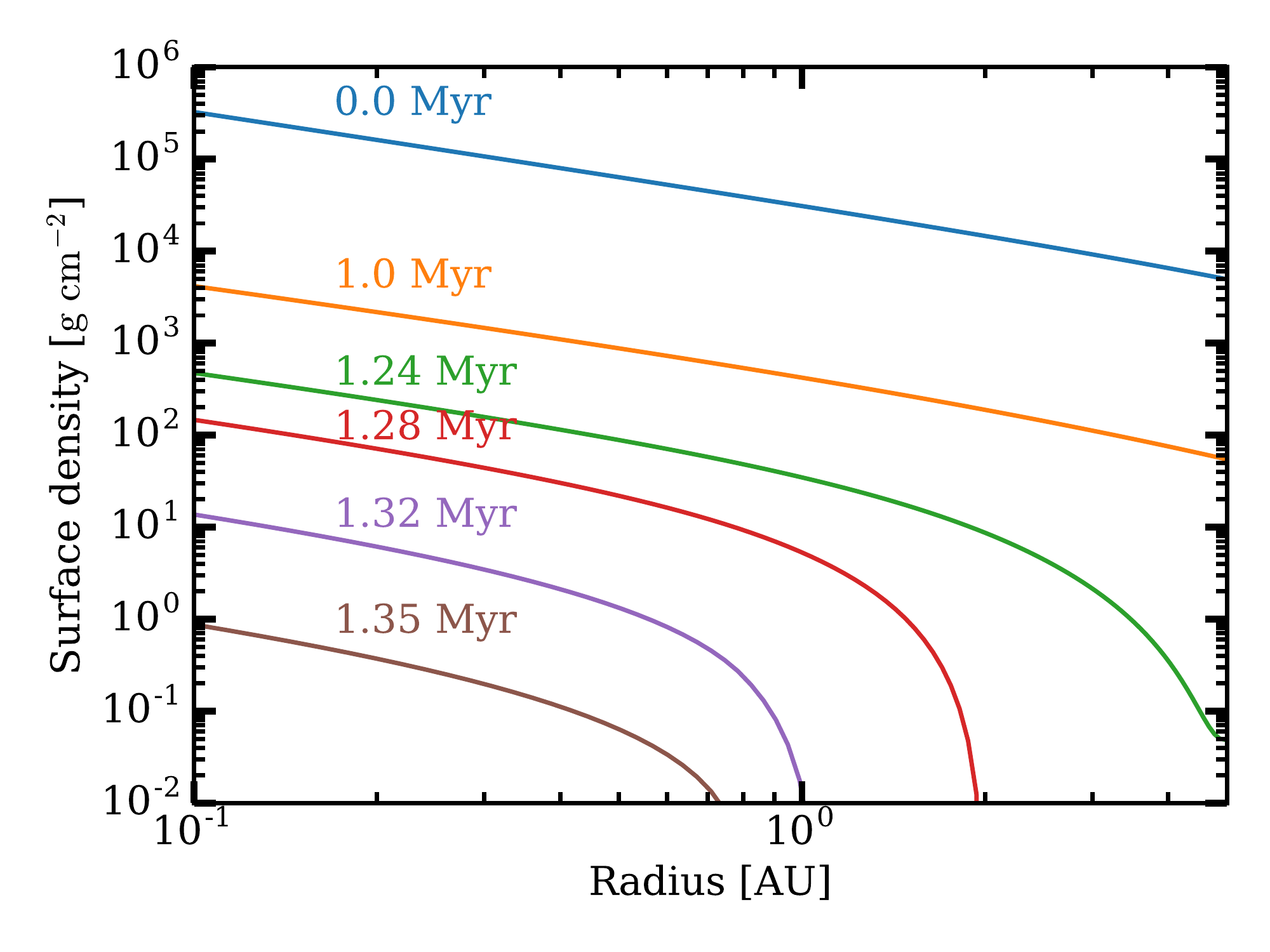}
\caption{The main patterns of photoevaporative disc clearing
in binaries identified in this paper. We plot snapshots of the surface density at selected times (indicated on the plot) for models with $M=1 M_\odot$ and $L_X = 2.3 \times 10^{30}$ erg s$^{-1}$. \textbf{Upper panel:} Tidal radius of $140$ AU. The clearing sequence
(with a gap forming at a few A.U.) is similar to that seen in discs
around single stars. \textbf{Mid panel:} an example of a disc
that does not form an inner hole in a model with a
tidal radius of $13$ A.U. The disc develops a gap and the outer part of the disc is removed by photo-evaporation. Subsequently the inner disc is cleared from outside in. \textbf{Lower panel:} a disc clearing purely outside in (without any gap) in a model with a tidal radius of $5$ AU.}
\label{fig:sigmat}
\end{figure}

 We find that the qualitative sequence of photoevaporative/viscous
clearance that we outlined above (a)-e)) is also followed in the binary case 
provided  $R_\mathrm{t}$
is significantly larger than the location of gap opening; in practice
this means cases where $R_\mathrm{t} > R_{crit} \sim 20 (M/M_\odot) $ A.U., although 
$R_{crit}$ can be 
as large as $100$ A.U. for the highest X-ray luminosities. 
The upper panel of Figure \ref{fig:sigmat} depicts the evolution of the surface density profile in such
a case. 

 For discs with $R_\mathrm{t} < R_{crit}$ (see above), an inner hole never
forms. The mid panel of figure \ref{fig:sigmat} shows the time evolution of such a disc. While a gap does open in the disc, the mass that would normally be in the region between the gap and the outer radius of the disc contains little mass and photo-evaporation removes it before the inner disc drains onto the star. In other words, phase (e) goes to completion before phase (d). Once photo-evaporation has removed the outer disc, the subsequent clearing of the inner disc proceeds from outside in.

For even smaller $R_\mathrm{t}$ ($\lesssim$ a few AU), the location of gap opening is outside the disc and therefore a gap never opens (see lower panel of Figure \ref{fig:sigmat}). In this case the disc clears purely from outside in. 

\begin{figure}
\includegraphics[width=\columnwidth]{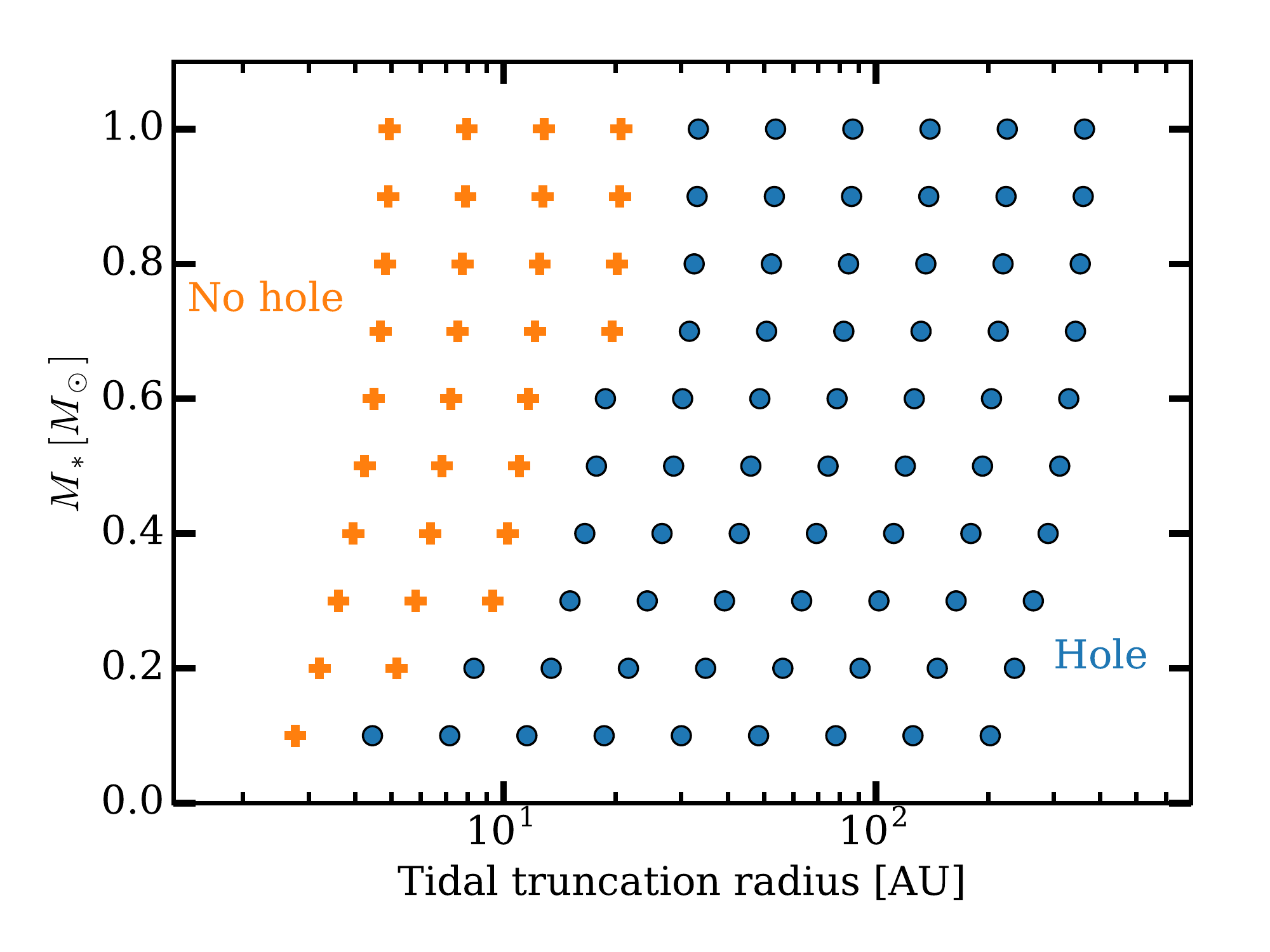}
\caption{The regions of parameter space (stellar mass versus tidal truncation radius)
for which inner holes form (blue circles) or not (green crosses) for average X-ray luminosity $L_X = 2.3 \times 10^{30} (M_\ast/M_\odot)^{1.44}$ erg s$^{-1}$.}
\label{fig:mass_rtrunc_hole}
\end{figure}

Figure \ref{fig:mass_rtrunc_hole} shows the range of parameters for which inner holes do
(blue dots) and do not (orange crosses) form for stars with average X-ray luminosity $L_X = 2.3 \times 10^{30} (M_\ast/M_\odot)^{1.44}$ erg s$^{-1}$. On the x axis we plot the tidal truncation radius of the disc according to the formula shown in figure \ref{fig:truncation}; the value depends on the combination of both the mass ratio and the binary separation. It can be seen how the $R_\mathrm{crit}$ dividing the discs that form holes from the ones that do not depends on the stellar mass, because of the radial scaling of the mass loss profile (see equation \ref{eq:radial_scaling}).

\subsubsection{Disc lifetime}
 \label{sec:disc_lifetime}

 We have shown that, if
$R_\mathrm{t} > R_{crit}$, photoevaporation creates inner holes
(just as in single stars) and these  may be  identified
with (at least some) observed transition discs \citep{Owen2012,OwenClarke2012}. This does not mean however that the effect of the companion is completely negligible, and both the \textit{absolute} lifetime\footnote{`Lifetime' is here
defined as the time taken for the disc to clear out to the smaller
of $R_\mathrm{t}$ and $100$ A.U.} of the disc and the \textit{fractional} time spent with an inner hole (`transition disc') can be modified.

In single stars, the lifetime of a disc against photoevaporation
is set by the time that is required for the viscous accretion rate
in the disc to decline to the level of the photoevaporative mass loss
rate. According to the viscous similarity solution (equation \ref{eq:self_similar_t}), the accretion rate in a freely expanding disc
declines as
\begin{equation}
\dot M = \dot M_{in} (1 + t/t_{\nu 1})^{-1.5},
\end{equation}
where $t_{\nu 1}$ is the viscous timescale at $R_1$ and
$\dot M_{in}$ is the initial accretion rate through the disc (which is 
proportional to the initial disc mass divided by $t_{\nu 1}$). 
In practice  the disc lifetime is thus set by the initial mass of the disc, its viscous evolution timescale, and the rate of
photoevaporation. In the case of a disc in a binary, the value of
$R_\mathrm{t}$ (which is a function of $q$ and $a$) is also relevant (see
\citealt{Armitage1999}). As discussed previously, after a time $t_\mathrm{boundary}$ that depends on the initial conditions and on the magnitude of the viscosity, the disc spreads enough to interact with the outer boundary, and from that time on according to equation \ref{eq:closed_boundary_sigma} the mass accretion will decrease as
\begin{equation}
\dot{M}(t) = \dot{M}(t_1) \exp \left( -\frac{3 \nu_0 \pi^2 t}{16 R_\mathrm{t}^2}\right)
\label{eq:mdot_exp}
\end{equation}
which decays in an exponential fashion with time, with a time scale that is the viscous time scale  $t_t$ at the outer edge. Therefore, even in the absence of photo-evaporation, discs with a small enough outer radius will clear\footnote{This does not take into account the initial expansion up to $R_\mathrm{t}$. In practice, we note that this happens on a timescale smaller or comparable to $t_t$ and thus does not change significantly the estimate} in $2-3 \ t_t$. The presence of photo-evaporation further hastens disc dispersal. We thus expect the disc lifetime to depend strongly on the value of $R_\mathrm{t}$.

\begin{figure*}
\includegraphics[width=\columnwidth]{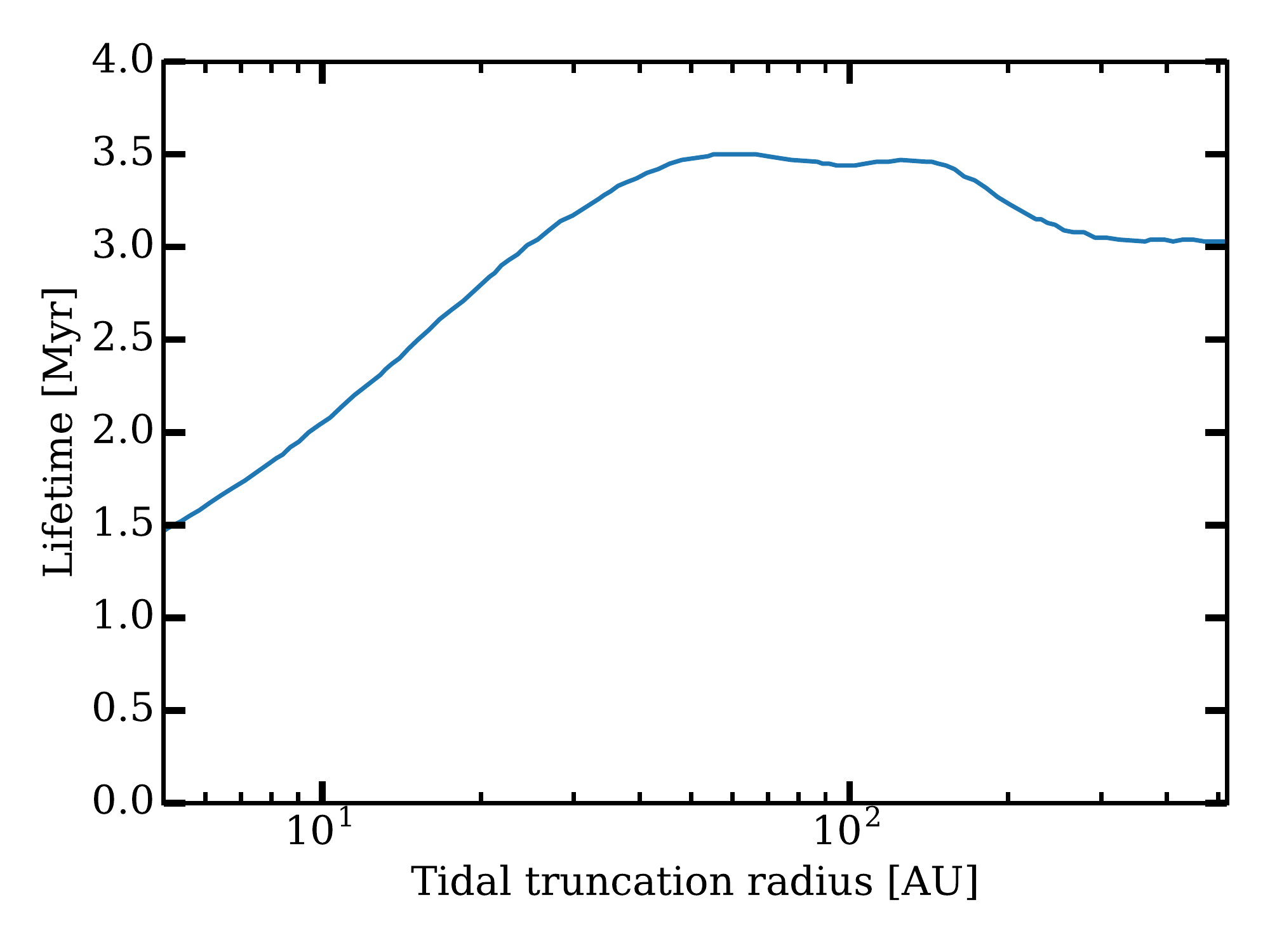}
\includegraphics[width=\columnwidth]{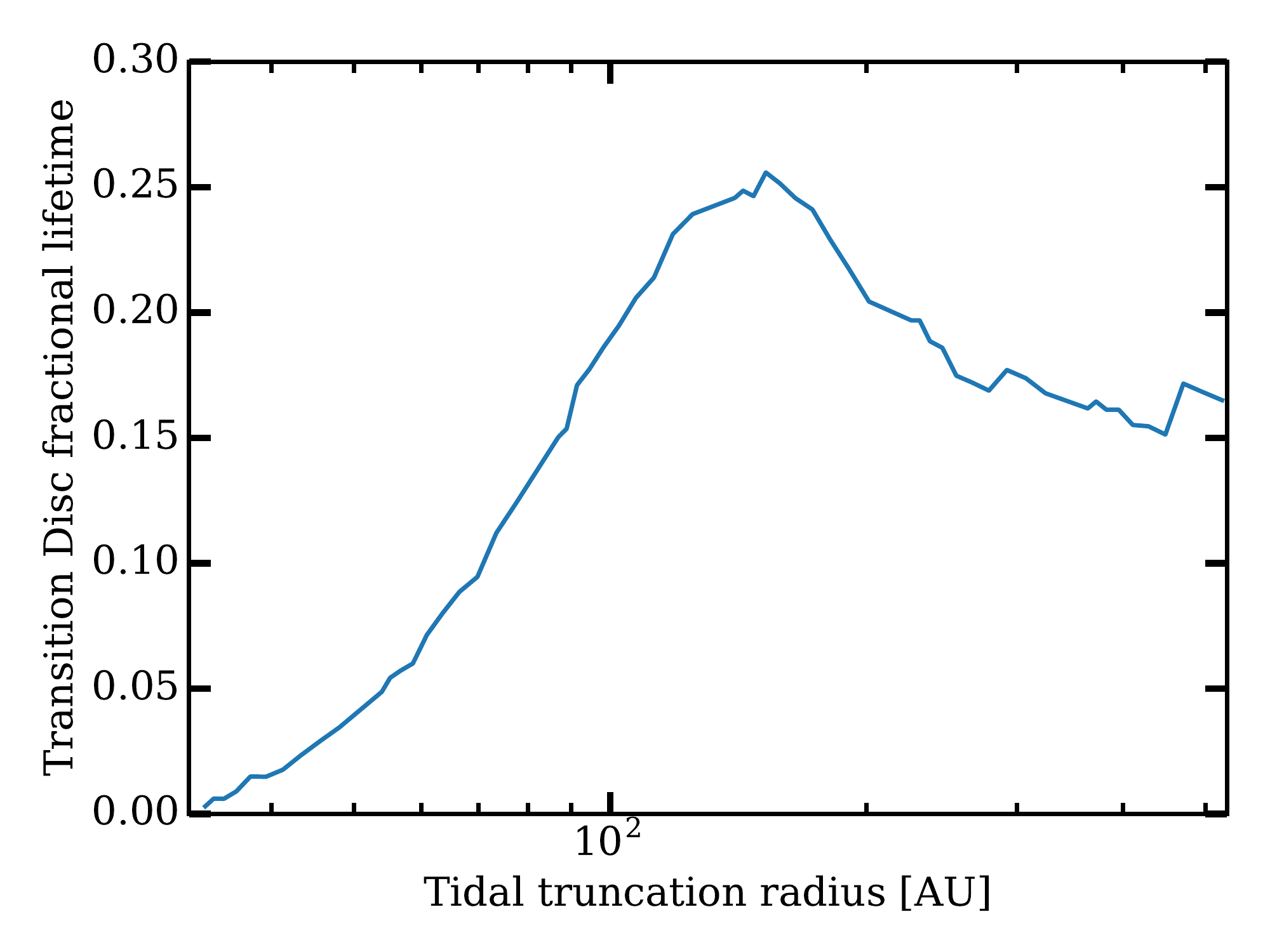}
\caption{Lifetimes as a function of the disc outer radius, set by the tidal forces of the companion. Both cases are for a star with $M=1 M_\odot$ with average X-ray luminosity ($L_X =
2.3 \times 10^{30}$ erg s$^{-1}$). \textbf{Left panel:} total lifetime of the disc. \textbf{Right panel:} fraction of the lifetime spent as a transition disc}
\label{fig:lifetime}
\end{figure*}

This expectation is borne out by Figure \ref{fig:lifetime}, which illustrates the dependence of the lifetime on the outer disc radius. Indeed, while for large outer radii ($ \gtrsim 100$  A.U.) there is little difference compared to a single star, for smaller radii the lifetime decreases dramatically. Figure \ref{fig:lifetime} shows also the dependence of the relative time spent as a transition disc (i.e., phases d-e). For radii smaller than $\sim 100$ AU, the fraction decreases and eventually vanishes for radii of $\sim 40$ AU, at which point the qualitative behaviour of the evolution switches to outside-in clearing (see previous section).
The reason for the reduction in the fraction spent as a transition disc is that after gap opening the bulk of the
remaining disc is retained in the region where the photoevaporation
is concentrated and also that the disc in this
region follows the steeper ($ \Sigma \propto R^{-1.5}$)  profile imposed
by the binary boundary condition, and therefore less mass is present outside the gap. 
Thus although inner holes can form in binaries, their incidence should be lower than
in single stars for outer radii smaller than 100 AU, corresponding to binary separations less than $\sim 300$ AU. Observationally, this means that we expect a smaller fraction of transition discs in binaries with a separation smaller than 300 AU. {The fraction of transition discs in binaries is currently poorly quantified, but it is interesting to note that the transition disc catalogue of \citet{vanderMarel2016} reports only 4 transition discs in binaries\footnote{We have excluded objects that are compatible with being circum-binary discs since in this paper we only study discs around each of the two components. Our final list comprises the following objects: 2MASS J04303235+3536133, 2MASS J04304004+3542101,  
2MASS J04292165+2701259, 	2MASS J16274028-2422040.}, out of a sample of $\sim 150$ objects.}


\subsubsection{Summary}
 
While disc clearing in wide binaries ($\gtrsim 30$ AU) proceeds in a similar fashion to single stars, the lifetime of transition discs
is expected to decline somewhat for close enough separations ($\lesssim 100$ AU) as it
becomes easier for photoevaporation to rapidly remove the remnant outer
disc.  For closer binaries ($\lesssim 30$ AU) we do not expect photoevaporation to create a hole but instead 
to produce progressive clearance of the disc from the outside in.
We stress that these estimates are based on a particular assumption
about the radial dependence of the kinematic viscosity  ($\nu
\propto R$) and that the detailed predictions would be expected
to change in the case of a more realistic viscosity model.   

\subsection{The differential evolution of discs within binary star systems}
\label{sec:differential}

In this section we now turn to consider the \textit{differential} evolution of the discs around both components of the binary, highlighting in particular the dependence on the stellar mass. We have shown in the previous section that the disc lifetime is set by the initial mass of the disc
and its viscous evolution timescale, as well as by the rate of
photoevaporation. We have also shown the dependence on the disc outer radius $R_\mathrm{t}$, which is a function of $q$ and $a$. On these grounds, discs around
secondaries should be expected to clear more rapidly on average because their discs are smaller (cf. Figure \ref{fig:truncation}) ;
\citet{Armitage1999} noted that this should imply a significant
fraction of binaries in which only the primary was associated
with a disc (i.e. CW pairs in the notation of \citealt{Monin2007},
where C refers to the Classical T Tauri star status of
disc possessing stars, W designates Weak Line (discless) T Tauri
stars and the ordering refers to the primary and secondary
respectively). However, additional factors are involved when
photoevaporation is included: the mean X-ray luminosity (and
hence wind photoevaporation rate) scales as $M_*^{1.44}$ compared
with the assumed linear scaling of the initial disc mass with
stellar mass: on these
grounds, one would expect the disc of the secondary star to be
longer lived and might thus expect the preferential production
of WC pairs.  

 \begin{figure}
\includegraphics[width=\columnwidth]{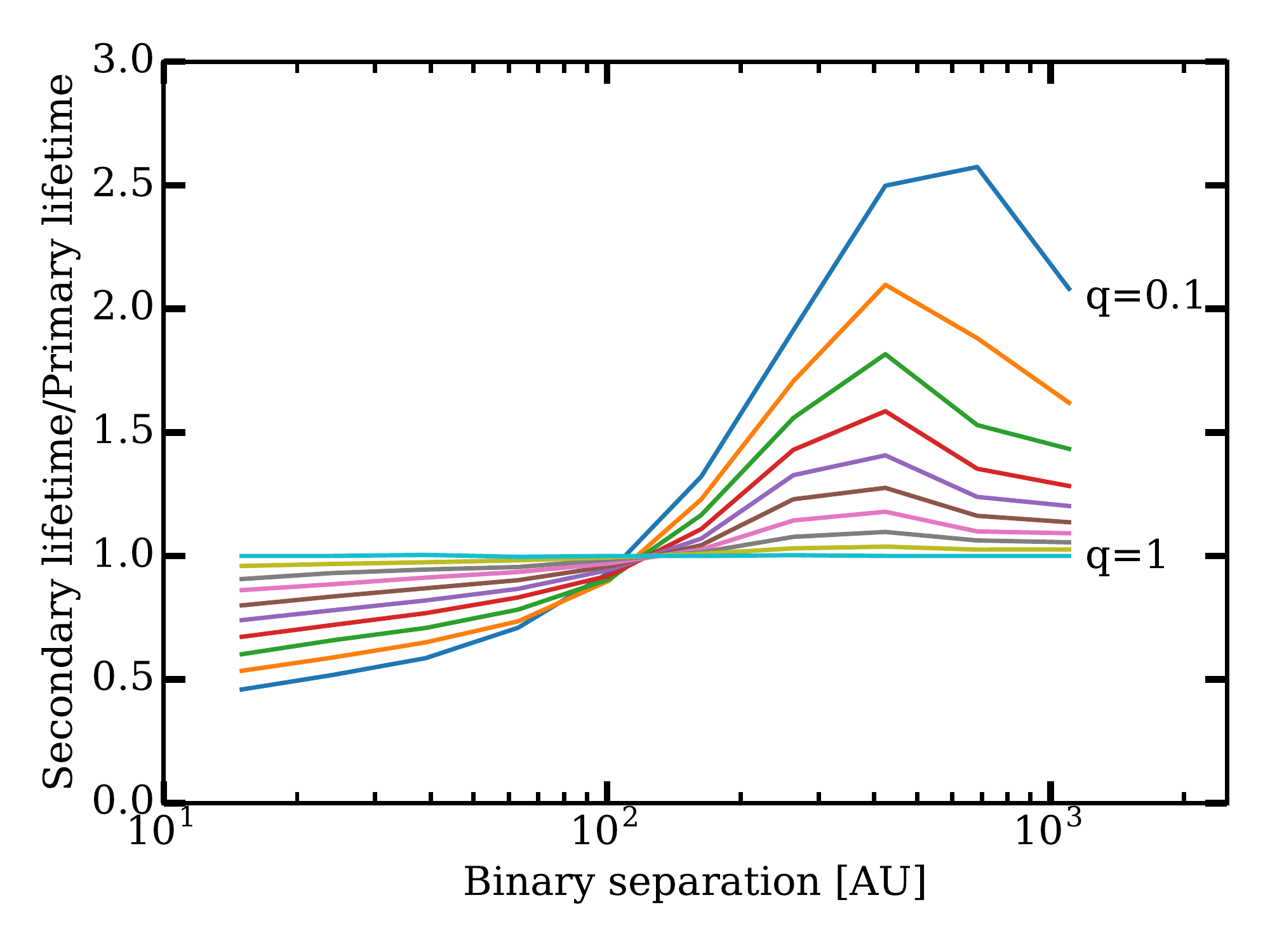}
\caption{The ratio of the lifetime of the secondary to that
of the primary (with a stellar mass of $1 \ M_\odot$) as a function of binary mass ratio and
separation, assuming that all stars follow the mean 
mass X-ray luminosity relation: $L_x = 2.3 \times 10^{30} (M_*/1 M_\odot)^{1.44}$ erg s$^{-1}$.
Discs around the secondary clear first for $a < 100 $ AU whereas
primaries clear first in wider binaries.}
\label{fig:differential}
\end{figure}

  Figure \ref{fig:differential} shows that both these effects come into play, depending
on the binary separation. For closer binaries, the disc is able to
viscously diffuse out to $R_\mathrm{t}$ before the discs are significantly
depleted by photoevaporation. Thereafter the accretion rate declines
exponentially (see equation \ref{eq:mdot_exp}), and it is thus the value of the e-folding time $t_t$
(related to the viscous timescale at $R_\mathrm{t}$) which determines the
time required for disc clearing, rather than the value of the
photoevaporation rate. Consequently secondary discs (with lower
viscous timecales at $R_\mathrm{t}$) clear somewhat more rapidly than
discs around primaries, even though they have lower
photoevaporation rates. At larger separations, the inverse is
true. The tidal boundary condition has less impact on the disc
evolution and even at the later evolutionary stages when
photoevaporation starts to become significant, the decline in disc
surface density and accretion rate is close to the
power law decline in time (equation 4)   
for a freely expanding disc. Consequently,
the level of the photoevaporative mass loss is important
in setting which disc clears first ( i.e.  the primary because of the
higher photoevaporation rate).

 At large separations, the ratio of 
 the secondary to primary
lifetime tends to a limiting value that can be readily understood
by considering the time at which the viscous accretion rate
equals the photoevaporation rate. The latter is given by
equation (4) so that (assuming that the initial disc mass
scales 
with the stellar mass)
we have $\dot M \propto M_* t^{-1.5}$. On the other hand
$\dot M_X \propto L_X^{1.14} \propto M_*^{1.65}$ and hence one expects
the disc lifetime to scale as $M_*^{-0.5}$. This is borne out by
Figure  \ref{fig:differential} (i.e. secondary lifetime roughly three times that of the
primary for $q = 0.1$ at the largest separations).   {In this work we assumed that the viscosity does not depend on the stellar mass. If the viscous time scales as $t_\nu \propto M_*^\beta$ (and hence the viscosity scales as $\nu \propto M_*^{-\beta}$, if the initial disc radius does not depend on the stellar mass as assumed in this paper), the scaling of the disc lifetime is modified as $M_*^{-0.5+\beta}$.}


 Figure \ref{fig:differential} represents the case that all stars have the average X-ray luminosity
for their masses. If we instead adopt X-ray luminosities that
are 1 $\sigma$ above ( below)  this average at all masses then the binary
separation at which the primary and secondary lifetime is equal shifts
to 70 AU and 240 AU respectively. This can be understood inasmuch as
differential photoevaporation effects become more (less) important at higher (lower)
X-ray luminosities.

\section{Discussion and comparison with the observations}
\label{sec:discussion}
We have investigated in the previous section how a disc in a binary evolves under the influence of viscosity and X-ray photo-evaporation and now proceed to
compare  our model predictions with observations of young binaries.
Our predictions of differential disc lifetimes need to be compared
with \textit{resolved} observations which can designate which member of the
binary pair possesses a disc (section \ref{sec:resolved}). As in section \ref{sec:differential}, in what follows we call systems in which both stars have discs
as CCs, and those with a disc around the primary (secondary) only
as CW (WC) systems. The reliable attribution of disc signatures to one
component or the other is however observationally challenging, particularly for small
separations and low mass ratios. We therefore also make use (in section \ref{sec:unresolved}) of the much
larger dataset of \textit{unresolved} observations in which the presence of
a disc cannot be attributed to a particular component but where
it is only possible to distinguish the three classes above from
a doubly disc-less (WW) system.


\subsection{The reduced lifetimes of discs in binaries}
\label{sec:unresolved}

In this section we consider \textit{unresolved} studies.
\citet{Cieza2009} first reported a statistical difference in the separations of binaries with and without discs. Binaries without discs tend to have smaller separations, which can be explained if the disc lifetime is reduced in tight binaries. This initial finding has been confirmed by \citet{Kraus2011}, who has shown in a sample of binaries in the Taurus-Auriga star forming region that the disc fraction is a function of the binary separation. Tight binaries ($a < 40$ AU) tend to have a smaller disc fraction than wide binaries, which instead have a disc fraction very similar to singles ($\sim 80$ \%). From the arguments presented in Section \ref{sec:results} we expect this result since we have shown that the disc lifetime decreases with decreasing binary separation (see Figure \ref{fig:lifetime}). 

We can test this \textit{quantitatively} by undertaking a population synthesis of discs. The goal is not to ``fit'' the observations, but rather to test whether the same model that is able to reproduce the evolution of the disc fraction in singles can reproduce also the observations in binaries. We now assign to each primary a mass according to the \citet{Kroupa2001} initial mass function (IMF). We then assign a mass to the secondary assuming a flat distribution in mass ratios, as found both in the solar neighbourhood \citep{Raghavan2010} and in the Taurus star forming region \citep{Daemgen2015}. We use the study of \citet{Raghavan2010} also to assign the separation, assuming a log-normal distribution in periods with an average of $10^5$ days and a standard deviation of $2.28$. We assign randomly X-ray luminosities to each star assuming a log-normal distribution with mean value $10^{30.37}$ erg s$^{-1}$ and a scatter $\sigma=0.65$, which is compatible with the observations in Orion of \citet{Guedel2007}. We then scale the luminosity with $M_\ast^{1.44}$. For each model binary
system we designate the disc lifetime as the maximum of
the lifetime of the primary's and secondary's disc.

\begin{figure}
\includegraphics[width=\columnwidth]{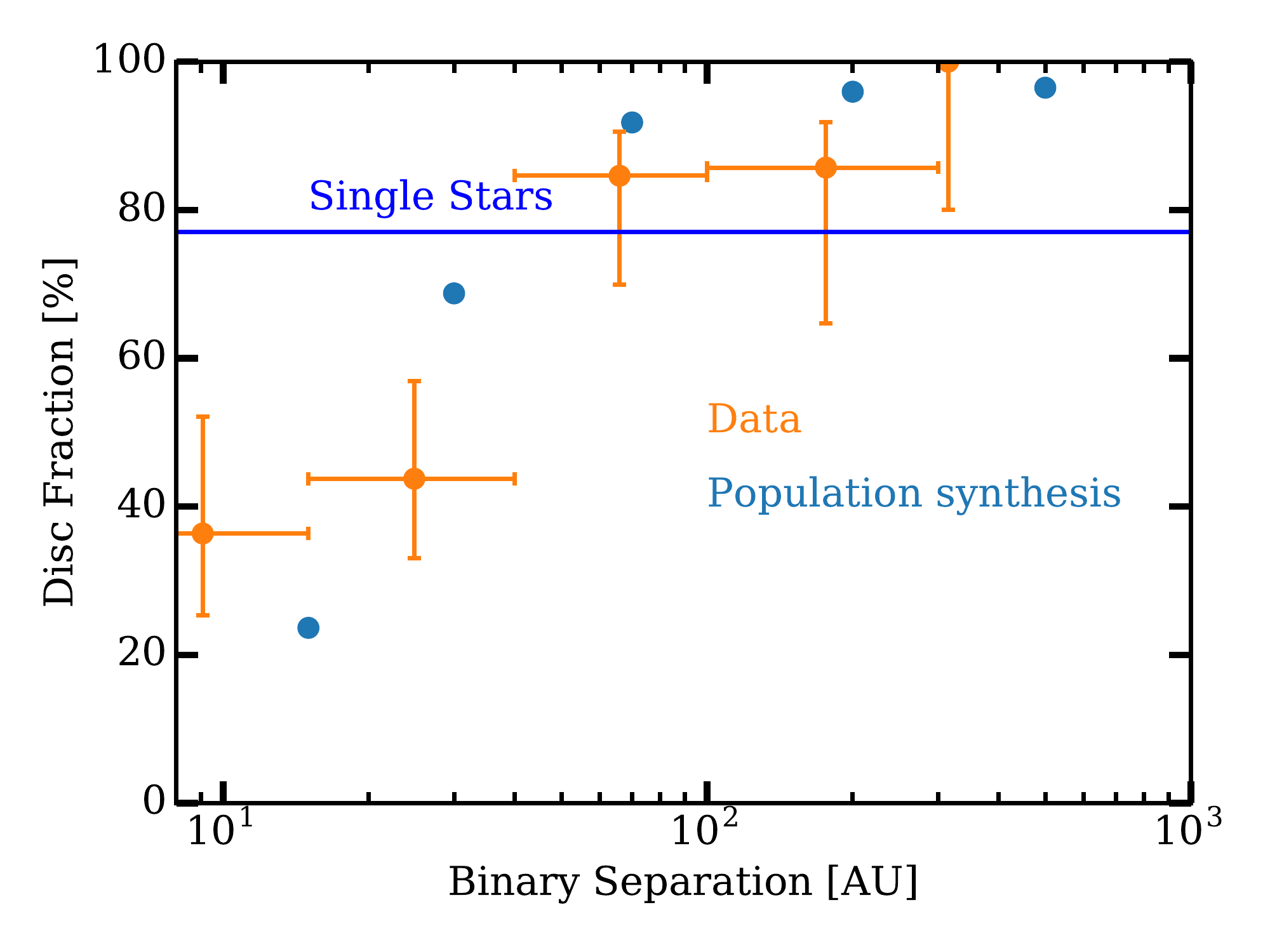}
\caption{The fraction of binaries {with a disc} after 2 Myr in our population synthesis (blue points) as a function of the binary separations. For comparison we plot also the data from \citet{Kraus2011} as the orange points with error bars. Wide binaries have a disc fraction slightly higher than singles (because the disc around the secondary is longer lived), whereas in tight binaries (i.e. $a \lesssim 50$ AU) the disc fraction is significantly reduced due to the shorter viscous time.}
\label{fig:fractionbin_a}
\end{figure}

In Figure \ref{fig:fractionbin_a} we plot the resulting disc fraction for the binaries as a function of separation {(blue points)}. To compute the fractions we have assumed an age of Taurus of 2 Myr and binned the systems by separation. For comparison we plot also as the horizontal line the disc fraction in singles from our models, which we have computed with a population synthesis with the same parameters (except that we do not use a closed boundary condition). The figure shows that wide binaries have a disc fraction similar to singles, although because the disc around the secondary for these separations is longer lived (Figure \ref{fig:differential}) the disc fraction is actually slightly higher than in singles. Notice that this effect is not visible in the observational data of \citet{Kraus2011}, but it is not incompatible with this interpretation. For tight binaries, the disc fraction is reduced. {As discussed in section \ref{sec:disc_lifetime}, this effect is mainly driven by the reduced viscous time at the outer boundary of the disc.} The model predicts that the reduction {in disc lifetime} happens for a turnover separation of $\sim 50$ AU, which is consistent with the observations. For the tightest binaries ($a \sim 10$ AU), the disc fraction reaches 10\%, which is less than the fraction of 25 \% measured by \citet{Kraus2011}. However, we notice that for these separation it is likely that the approximation that there is no circumbinary disc (which is longer lived for close separations, \citealt{AlexanderBinary2012}) is probably no longer adequate. Moreover, the low number statistics of the observations means that there is considerable uncertainty in the measured fractions. 

{It is interesting to note that, although the reduced disc lifetime at close separation is \textit{mainly} a viscous effect, the inclusion of photo-evaporation is still required to hasten further disc dispersal. To show this, we have run the same calculation again without including photo-evaporation. We find that in this case no disc is completely dispersed at the age of Taurus. This might perhaps be cured by shortening the viscous time. As already stated, however, it is not our goal here to fit the observations performing a full investigation of the degeneracies between the parameters. We conclude that the model we have \textit{assumed} as driver of disc evolution, where viscosity and X-ray photo-evaporation are the driver of disc evolution, is broadly consistent with these observations.}

\subsection{Differential disc evolution in T Tauri binaries} 
\label{sec:resolved} 
 
\begin{figure*}
\includegraphics[width=\columnwidth]{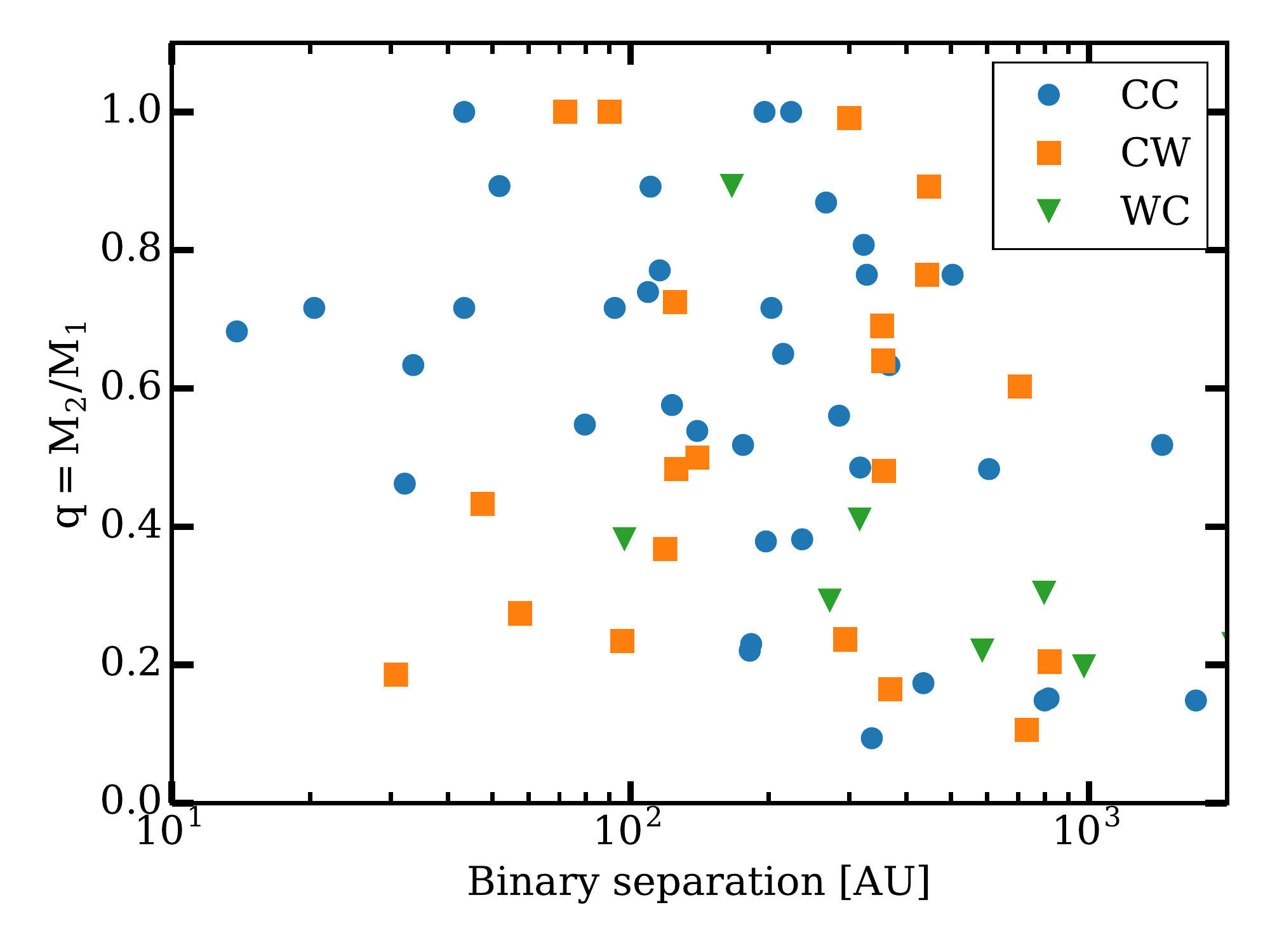}
\includegraphics[width=\columnwidth]{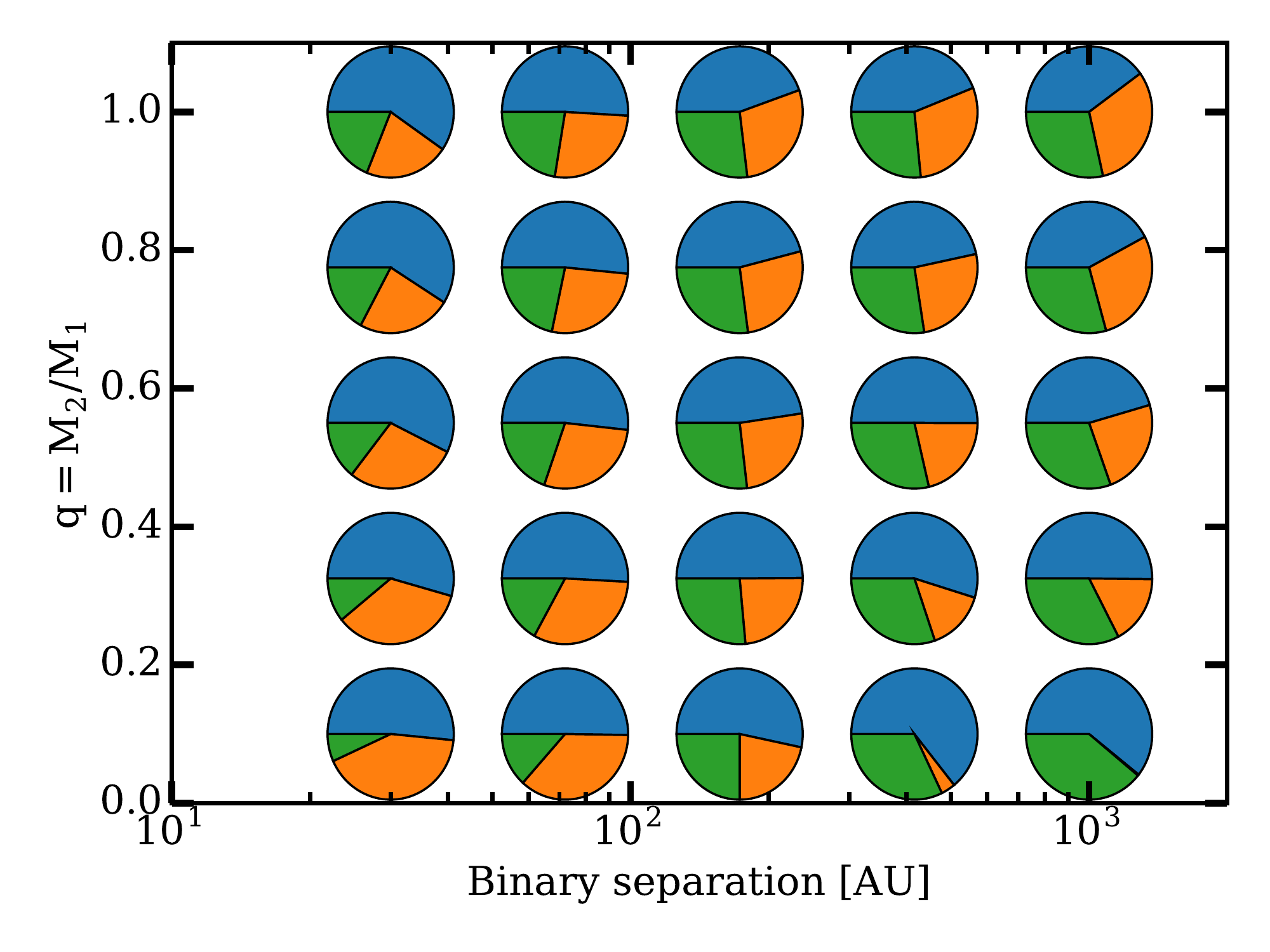}
\caption{Comparison between observations and our models for the \textit{differential} disc evolution in binaries. \textbf{Left panel}: compilation of results from \textit{resolved} observations. Blue circles are for CC systems, orange squares CW and green triangles WC. To place points in the plot we have computed the mass ratio from the published spectral types using the spectral type-effective temperature scale of \citet{HillenbrandWhite2004} and the pre-main sequence tracks of \citet{Siess2000} (assuming an age of 1 Myr). \textbf{Right panel:} A montage of results from the Monte Carlo simulations
showing pie charts of the relative times that the system spends
as a CC, a CW or a WC as a function of binary separation (colours have the same meaning as in the left panel: blue for CC, orange for CW and green for WC). Each pie
chart contains objects with a range of primary masses. The plots
illustrate that few WCs are expected at small separations while their incidence increases at large radii. For most of the parameter space a mixture of
CWs and CCs is expected. 
}
\label{fig:ccw}
\end{figure*}

In this section we consider \textit{resolved} studies. We have collected data from the literature using the catalogue of \citet{Monin2007} \citep{ReipurthZinnecker1993,Kohler2000,Hearty2000,Koresko2002,Correia2006} and the new observations that have been taken since then \citep{Comeron2009,Daemgen2012,Daemgen2013}. We have plotted the distribution of observed CCs, CWs and WCs in the plane
of separation versus mass ratio in the left panel of Figure \ref{fig:ccw}. To compute the mass ratios, we have converted spectral types to masses using the spectral type-effective temperature scale of \citet{HillenbrandWhite2004} and the pre-main sequence tracks of \citet{Siess2000} (assuming an age of 1 Myr).
We do not plot WWs because the
number of WWs relative to the other categories depends on the 
total length of time that stars spend as T Tauri stars, whereas
the expected ratio of CWs and WCs to CCs can be directly inferred from
the ratio of lifetimes. 

In order to assess whether the observational results are compatible with the observations, we need to recall that we do not have information
on the individual X-ray luminosities of the components in the
observed binaries since the X-ray observations can only disentangle
pairs separated by $\sim 1000$ A.U. \citep{Preibisch2005,Guedel2007}. Moreover, the numbers of objects observed is insufficient
to be able to group the data as a function of primary mass, so that
we cannot directly use a plot like Figure \ref{fig:differential} (which has a fixed primary
mass and assumes that the X-ray luminosities of each component
follow the mean X-ray luminosity as a function of stellar mass
relationship). Instead, we here undertake population synthesis
experiments: binary primaries are selected according to the
IMF of \citet{Kroupa2001}, while X-ray luminosities are attached
to each component by selecting from the same X-ray luminosity function used in the previous section. This is an important difference from the calculations
reported in Section \ref{sec:results} because it is now possible, for example,
for a secondary star to have a significantly higher X-ray luminosity
than a primary star.

The model allows us to compute the expected fraction of systems that are CC, CW or WC. This fraction is a function of time (after enough time, all systems become WW); however, by assuming that we are sampling uniformly the distribution of stellar ages bearing discs, we can compute a time-averaged value (the situation is the same described in \citealt{Rosotti2015} for the fraction of transition discs). This assumption is justified because the data with which we are comparing is a compilation of observations from different regions of different age. 



In the right panel of Figure \ref{fig:ccw} we show a montage (as a function of binary separation
and mass ratio) of the resulting fractions of systems which would be expected to be in the CC, CW and
WC category. The colours have the same meaning as in the left panel. At the closest separations, WCs are expected to be
rare, particularly at low mass ratios (i.e. high $M_{\rm primary}/M_{\rm secondary}$): as discussed in Section \ref{sec:results}, the differential evolution in tight
binaries is driven mainly by the difference in viscous evolution
timescale and hardly at all by the differential levels
of X-ray photoevaporation. For all mass ratios not equal to
unity, the ratio of WCs to CWs increases with binary separation. 
The scatter in X-ray luminosities is important here. Whereas Figure
\ref{fig:differential} (which assumes the mean mass-Xray luminosity relationship) suggests that {\it all} binaries at a separation of a few hundred A.U. should pass through
a WC, rather than CW, phase, Figure \ref{fig:ccw} shows that this is not
the case: only for the most extreme mass ratios at large separations (lower right corner of the plot) no CW system is predicted, and for the regions of parameter spaces where the majority of mixed pairs
are observed, the model predicts comparable numbers of CWs and WCs. This is not found in the observations, which find only a very limited number of WC systems. It is however likely that there is a selection bias against undertaking resolved studies on systems that turn out to be WCs: a faint excess around the secondary could be overlooked in unresolved
studies, especially at small separations and low mass ratios. Although we are
unable to quantify this effect, we note that those WCs that have been observed
tend to occupy the lower right area of parameter space which is where the model
predicts that they should be more predominant.


The model also predicts a comparable abundance of CWs and CCs over most of the parameter space. This is compatible with the observations: in general no systematic difference between the blue and orange points can be observed. There are however two notable exceptions. The first is for systems with very small separations ($a \sim 30$ AU) and mass ratios $q \gtrsim 0.4$. In this case, the model predicts an increase in the fraction of CCs, which seems to be confirmed by the relative lack of orange points in that region.  The second concerns systems at large separations and small mass ratios, where the model predicts that there should be no CW system due to the much longer lifetime of secondaries. There are two CW datapoints with a separation of $\sim$800 AU and a mass ratio below 0.2 that cannot be reconciled with the model. This could mean that {the actual dependence of the disc lifetime with stellar mass is shallower than the one predicted by the model we employed. As discussed in section \ref{sec:differential}, this is a consequence of the scaling of the viscous time with stellar mass that we assumed; a shallower scaling of the disc lifetime could be obtained by assuming that the viscous time mildly increases with the stellar mass, i.e., using the notation of section \ref{sec:differential}, $0<\beta<0.43$.} Given the small number of incompatible points, we will not discuss further this discrepancy. We stress however that the discrepancy highlights that this region of the parameter space with large separations and small mass ratio is ideal to study the dependence of the disc lifetime with the stellar mass, but the limited size of the current sample hinders further progress on the subject.

We conclude that the predictions of the model are compatible with the data available at the moment, provided that there is a significant bias against the detection of WC systems.


\subsubsection{The dependence of disc lifetime on stellar mass}
{In our model, discs around isolated low-mass stars or low mass stars in wide binaries are longer lived, which is the reason for the increased fraction of WC systems in the bottom right corner of Fig. \ref{fig:ccw}. We have shown in this paper that this is compatible with the observations. The dependence of the disc lifetime on the stellar mass has many implications for understanding what kind of planetary systems may form around these stars. There is some indication from Kepler that the ratio of mass contained in planets
to stellar mass is higher in low mass stars \citep{Mulders2015} and an increased disc lifetime might provide an explanation for that.}

{We note that the dependence of disc lifetime with stellar mass has been studied in several other works. Most of them have used disc fractions deduced from the infra-red (typically using Spitzer), subdividing stars in mass bins and then studying in which bin the disc fraction is lower (which is interpreted as evidence that discs are shorter-lived). We note that this has produced contrasting results; while some studies have found that discs around low-mass stars are longer-lived \citep[e.g.,][]{Carpenter2006,Kennedy2009,Ribas2015}, the opposite result has also been claimed in the literature \citep{Luhman2008,Luhman2010}. The differences can be probably ascribed to the different environmental conditions of the regions studied, as well as the difficulties in reaching completeness, which is essential for measuring disc fractions (this problem is particularly severe for low mass, disc-less stars). With a different method, \citet{ErcolanoBastian2011} instead reached the conclusion that there is no systematic difference in the disc lifetime between low and high mass stars since there is no systematic difference between their spatial distribution. In this respect, we note that a possible explanation of their result is if the age spread in the region is lower than the disc lifetime. Finally, \citet{Kastner2016} recently analysed the stars in the TW Hya association, finding that a significant fraction of the stars later than mid-M still retain their disc, in contrast to the earlier spectral types. In addition, they found a decline in the ratio of X-ray luminosity to bolometric luminosity at late spectral types. Both these observational pieces of evidence are in agreement with our model.}

{We point out that binaries with wide orbits and low mass ratios offer an additional way to study the dependence of the disc lifetime on the stellar mass. While the current constraints are limited due to the small size of the current sample of binaries, we note that binaries on wide orbits are observationally relatively easy to access. The fact that the primary is intrinsically brighter also mitigates the problem of finding a significant sample of low mass stars. This is therefore a promising avenue for further studies.}

\subsection{Combining mass accretion rates and mass measurements}
So far when comparing with observations we have only considered if there is a disc or not, either in general in the binary (\textit{unresolved} studies), or around a specific component (\textit{resolved} studies). Observations are starting to provide more information, by measuring the mass of each individual disc \citep{Harris2012,Akeson2014} and the accretion rate on each star \citep{Daemgen2012,Daemgen2013}. The additional information available places more constraints on the evolution of these discs; for example, resolved observations in the sub-mm have demonstrated \citep{Jensen96,Harris2012} that discs in binary systems have a lower mass than discs around single stars.

\begin{figure}
\includegraphics[width=\columnwidth]{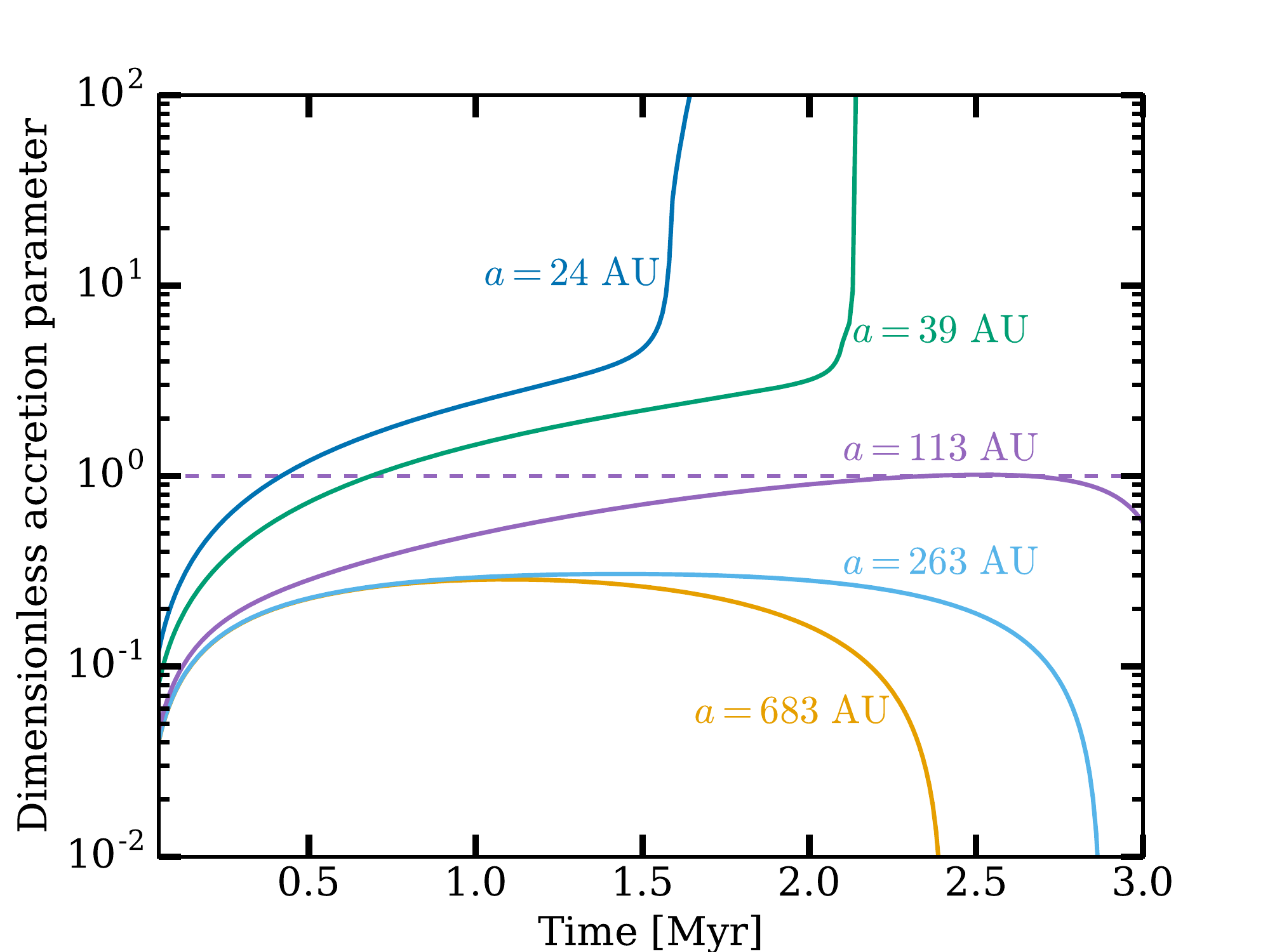}
\caption{The evolution of the dimensionless accretion parameter in binaries with different separations (indicated on the plot). For illustrative purposes we have chosen equal mass binaries with a mass of $1 M_\odot$ with average X-ray luminosity. In wide binaries the parameter becomes significantly lower than unity due to the opening of a hole and the suppression of accretion onto the star; in close binaries the parameter is higher at all times, a factor of a few throughout most of the evolution and then increasing significantly during a short burst of accretion at the end as the disc is cleared from outside in.}
\label{fig:mmdot}
\end{figure}

\citet{Rosotti2017} (see also \citealt{Jones2012}) have shown how measurements of disc masses and accretion rates can be combined through the dimensionless accretion parameter $\eta = t \dot{M}/M_\mathrm{disc}$, where $t$ is the age of the system. Figure \ref{fig:mmdot} shows the evolution in time of $\eta$ in a binary system. For illustrative purposes we have chosen equal mass binaries with a mass of $1 M_\odot$ with average X-ray luminosity and a range of separations indicated on the plot. However the qualitative behaviour we describe is general. The figure shows that for wide binaries (corresponding to the ones that form a hole, see section \ref{sec:qualitative}) the evolution of $\eta$ follows that expected for single stars that are eventually cleared by photoevaporation, i.e. $\eta$ is of order unity for most of its evolution but declines steeply at late times during inside-out clearing. In close binaries the accretion parameter instead increases monotonically with time, with a steep increase at late times during outside-in
clearing.  This behaviour is similar to the one in discs clearing under the effect of external photo-evaporation. Thus, for the \textit{same} disc mass, we expect close binaries to have a higher accretion rate (for most of the disc lifetime, of a factor 3-4).

\begin{figure}
\includegraphics[width=\columnwidth]{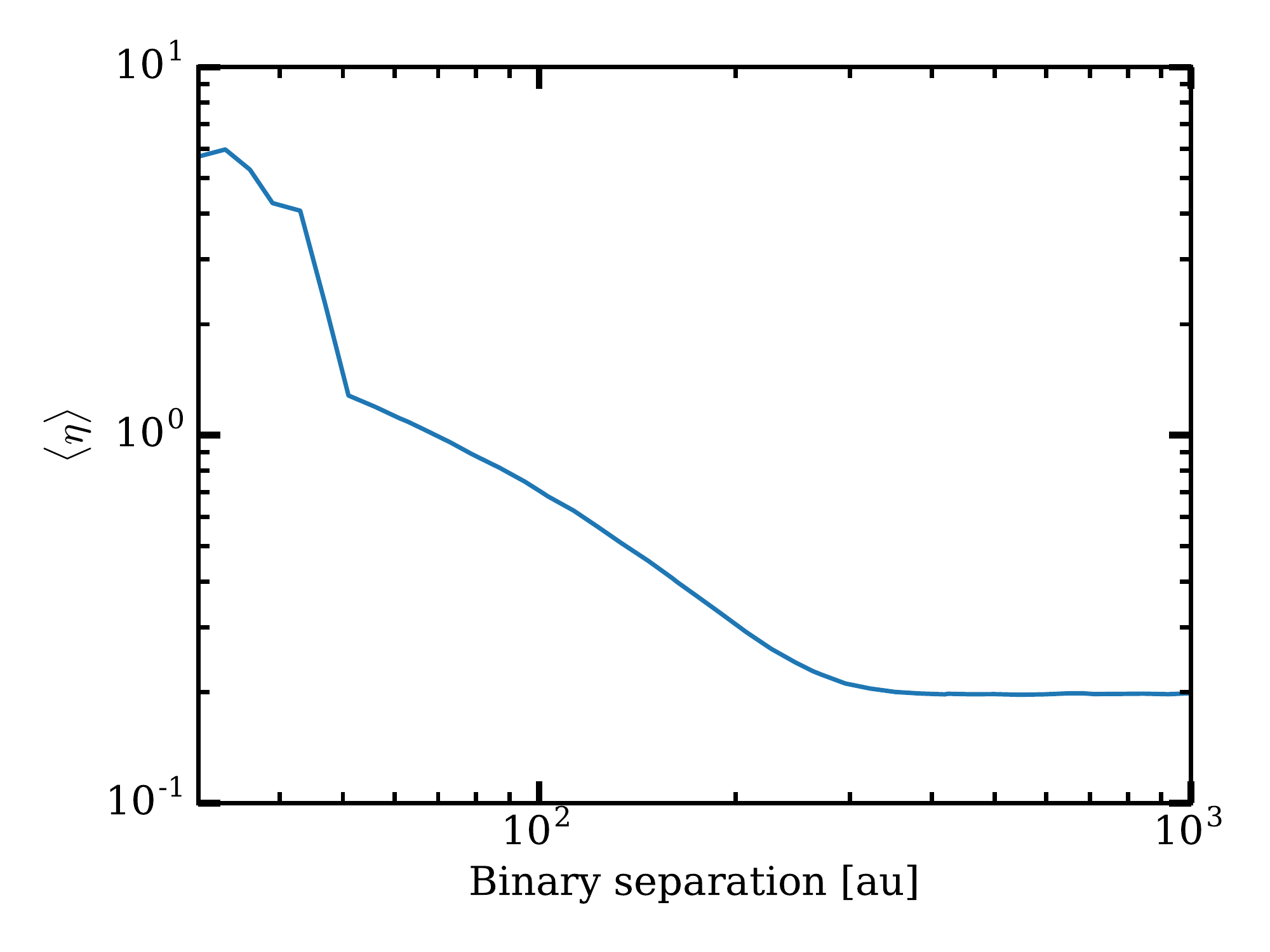}
\caption{The dimensionless accretion parameter $\eta$, averaged over the disc lifetime, as a function of the binary separation.}
\label{fig:eta}
\end{figure}

To be more quantitative about the predicted change in the value of $\eta$ in binaries, in Figure \ref{fig:eta} we plot $\eta$ as a function of the binary separation. Because $\eta$ is a function of time, we have averaged $\eta$ over the disc lifetime. Observations sample disc at different phases of evolution and the time average can thus be compared with the distribution of $\eta$ yielded by observations. The figure shows that there are three different regimes for $\eta$ depending on the binary separation. The flat part of the curve at large separations ($a \gtrsim 300$ au) corresponds to discs that are dispersed before the outer radius has reached the tidal truncation radius. Disc evolution in these binaries proceeds as in single stars. Moving to closer separations, $\eta$ increases monotonically as the binary separation is reduced. Binaries with an intermediate separation 50 au $< a < 300$ au still clear from inside out (see section \ref{sec:mean}) because of photo-evaporation. However, because of the interaction with the outer boundary the accretion rate onto the star is enhanced, and thus $\eta$ is a factor of a few higher than in single stars (for an example of the time evolution of $\eta$, see the $a=113$ au case in figure \ref{fig:mmdot}). Finally, for very close binaries which experience outside-in clearing $\eta$ greatly increases (more than a factor of 10).


 To the best of our knowledge, there is currently no close binary system where both the disc mass and the accretion rate have been measured for individual components. Future observations will be able to test these theoretical predictions. There are already, however, observational implications as \textit{unknown} binaries are likely present in many currently studied samples of T Tauri stars. This will affect the determination of the disc mass and of the accretion rate onto the stars. \citet{Manara2016} have shown observationally in the Lupus star forming region the existence of a correlation between disc mass and accretion rate, which points to a constant value of $\eta$ (approximately unity), consistent with disc evolution under the effect of viscosity. There is however a significant scatter of $\sim$ 0.5 dex in the correlation. Here we point out that the presence of binaries in the sample contributes to this observed scatter. In the case of a low mass ratio $q \ll 1$, the primary will outshine the secondary so that the measured quantities (i.e., stellar mass, disc mass and mass accretion rate) will reflect the ones of the primary star. This will lead to an increase in the measured mass accretion rate with respect to a comparison sample of single stars. Contamination by binaries will therefore increase the scatter in the correlation. The same increase in scatter obviously happens for binaries where both stars contribute significantly to the flux and the impact of fitting their emission with a single stellar model is more difficult to quantify. We thus expect the correlation between disc mass and mass accretion rate to become tighter when binaries are removed from the sample.

\subsection{Model limitations}
{Binaries are known to have a wide distribution of eccentricities (e.g. \citealt{Raghavan2010}), while in our model we have assumed that they are are circular. Some discs in binaries are also observed to be misaligned to the binary orbital plane (e.g. \citealt{JensenAkeson2014}). These two effects would cause some modification
of the truncation radii but would not affect the conclusion
that the secondary's disc  is more truncated than the primary, 
with the possible exception of the most strongly misaligned systems (see e.g. \citealt{ArtymowiczLubow1994,Pichardo2005} for works that included the effects of eccentricity and \citealt{2015ApJ...800...96L} for the effects of inclination). Another effect that we have neglected in this paper is the possible truncation of the disc at the location of vertical resonances \citep{Lubow1981,Ogilvie2002}. However, the truncation happens only if the viscosity in the disc is low enough to be effectively in a so-called dead zone. Finally, probably the biggest uncertainty in our models is the mechanism responsible for driving accretion; while in this work we have assumed it is viscosity, recent work \citep[e.g.,][]{Suzuki2009,BaiStone2013,Fromang2013,Simon2013} is suggesting that winds might also play an important role by removing angular momentum from the disc. The effect of winds on disc evolution is however still poorly understood and for this reason we neglected it in this study. Even if accretion is driven by viscosity, there are still
many associated unknowns, of which the most important from the
perspective of this paper is its dependence on stellar mass.}
   
\section{Conclusions}
\label{sec:conclusions}

Our calculations have modelled disc clearing by X-ray photoevaporation
in a binary environment, capitalising on the fact that the X-ray properties
of young stars, the mass loss profiles predicted by X-ray photoevaporation
theory and the tidal truncation of discs within binary systems
are all well determined properties. The most poorly determined
quantity that enters our models is the viscosity law in the disc. Here
we simply prescribe the viscosity as a power law of radius. While
this allows a ready comparison with previous work, it should be borne
in mind that this phenomenological description may not be
a good approximation to the viscous properties of real discs.

 Our results should be contrasted with those of \citet{Armitage1999}
(viscous only calculations in binary systems) and \citet{Owen2011} (X-ray photoevaporation calculations in single stars). 
Our main results are:

\begin{itemize}

\item We have shown that the sequence of disc clearing depends on the
location of the disc's tidal truncation radius compared with the location
of maximum X-ray photoevaporation (see Figure \ref{fig:mass_rtrunc_hole}). Except for the closest binaries (separations of a few AU), a gap always opens as in the evolution around single stars. Inner holes form
if the tidal truncation radius is well outside the region of maximum photoevaporative mass loss (see Figure \ref{fig:sigmat}).
In this case material remains in a ring outside the gap which is slowly eroded by photo-evaporation while the inner disc drains. If however the disc is tidally truncated close to the
maximum of photoevaporation (i.e. at a few tens of AU) then the outer ring clears faster than the inner disc. Instead of forming an inner hole,
we expect such discs to clear from the outside in. A consequence of this is that we should expect fewer transition discs (at least, of the type created by photo-evaporation, see \citealt{OwenClarke2012}) in binaries.

\item We have shown that in close binaries both discs see their lifetime reduced due to the smaller viscous timescale. The quantitative predictions of the model compare well with the observations in the 2 Myr old Taurus star forming region \citep{Kraus2011}.

\item We have shown that in closer binaries, discs tend to survive
longer around the primary than the secondary component (see Figure \ref{fig:differential}) due to the faster viscous time-scale induced by a smaller tidal truncation radius.
The converse is true in wide binaries,
where the differential evolution is instead driven by the higher mean
photoevaporation rate in the case of the more massive (primary)
component (i.e. on average discs survive for longer around the
secondary). Since there is a scatter of X-ray luminosity at all
masses, we expect a combination of WC/CW systems\footnote{See Section \ref{sec:discussion} for a description of the classification C/W} for all separations, but with a preponderance of WC at large separations and of CW at small separations. Figure \ref{fig:ccw} confirms that most WCs systems are found at large separations. For the model to be compatible with the observations however we need to assume that there is a general, not yet quantified observational bias towards detecting WC systems (only few of which are known), justified by the difficulty of detecting infra-red excess around the secondary due to the higher luminosity of the primary. With this assumption the current observational
data on the statistics of CC versus mixed pairs is broadly consistent
with the predictions of the model. 

{\item Our model implies a longer lifetime for discs around low mass stars; we note that this is consistent with, but poorly constrained from, the current observations. This is important to understand the process of planet formation around these stars. There is some indication from Kepler data that the ratio of mass contained in planets
to stellar mass is higher in low mass stars \citep{Mulders2015} and the increased disc lifetime might provide an explanation. We point out that binaries on wide orbits and with low mass ratios offer a way to study the dependence of the disc lifetime on the stellar mass, as an alternative to using disc fractions as a function of primary mass in star forming regions. Binaries on wide orbits are observationally relatively easy to access, making this a promising avenue for further studies provided that the observational sample is expanded.}

\item Our population synthesis has allowed us to assess the relative lifetimes
of primary and secondary discs in the regime of small separations where
observational data is sparse (see Figure 7). For example, we find that
for binaries closer than $30$ A.U. and with $q < 0.5$, the primary's
disc lifetime exceeds that of the secondary in the great majority of cases,
a finding with implications for the relative incidence of planets around
primary and secondary stars in binaries.

\item Future observations will provide another test of the models by combining measurements of both the mass accretion rate and the disc mass in the individual components of binary systems. Theoretically, we expect the dimensionless accretion parameter $\eta = t \dot{M} /M$ \citep{Jones2012,Rosotti2017} to be a function of the binary separation, and in particular \textit{higher} in close binaries (because of the shorter viscous timescale imposed by the outer boundary) than in singles. Because \textit{unknown} binaries likely contaminate existing observational samples, this effect increases the scatter in, for example, the correlation between disc masses and accretion rates reported by \citet{Manara2016}. {The scatter in the correlation was recently analysed by \citet{Lodato2017} and \citet{Mulders2017}; both papers based their analysis on the fact that the magnitude of the observed scatter places constraints on the system age relative to the initial viscous time of the disc. A careful study of the sample to remove binaries would be needed to quantify how much unknown binaries contribute to this increased scatter.}

\end{itemize}


\section*{Acknowledgements}
{We thank to the referee for useful comments that improved the paper}. We thank Sophie Kneller for her early work on the subject and Lisa Prato for giving us access to her observational database of resolved binary observations. This work has been supported by the DISCSIM project, grant
agreement 341137 funded by the European Research Council under
ERC-2013-ADG, {and by the Munich Institute for Astro- and Particle Physics (MIAPP) of the DFG cluster of excellence "Origin and Structure of the Universe"}.

\bibliographystyle{mnras}
\bibliography{binclear}

\bsp
\label{lastpage}
\end{document}